\documentclass[
reprint,
superscriptaddress,
longbibliography,
amsmath,amssymb,
aps,
prl,
floatfix,
]{revtex4-2}

\usepackage{graphicx}%
\usepackage{xcolor}
\usepackage{cancel}
\usepackage{dcolumn}%
\usepackage{bm}%
\usepackage[colorlinks, linkcolor=blue, citecolor=blue, urlcolor=blue]{hyperref}%
\usepackage{booktabs}

\begin{document}

\title{Anyonic Mach-Zehnder interferometer on a single edge of a 2D electron gas}

\affiliation{Department of Physics, Brown University, Providence, Rhode Island 02912, USA}
\affiliation{Brown Theoretical Physics Center, Brown University, Providence, Rhode Island 02912, USA}
\affiliation{Department of Physics, University of Illinois at Urbana-Champaign, Urbana, IL, USA}

\author{Navketan Batra}
\affiliation{Department of Physics, Brown University, Providence, Rhode Island 02912, USA}
\affiliation{Brown Theoretical Physics Center, Brown University, Providence, Rhode Island 02912, USA}
\author{Zezhu Wei}
\affiliation{Department of Physics, Brown University, Providence, Rhode Island 02912, USA}
\affiliation{Brown Theoretical Physics Center, Brown University, Providence, Rhode Island 02912, USA}
\author{Smitha Vishveshwara}
\affiliation{Department of Physics, University of Illinois at Urbana-Champaign, Urbana, IL, USA}
\author{D. E. Feldman}
\affiliation{Department of Physics, Brown University, Providence, Rhode Island 02912, USA}
\affiliation{Brown Theoretical Physics Center, Brown University, Providence, Rhode Island 02912, USA}

\date{\today}

\begin{abstract}
Anyonic Fabry-P{\'e}rot and Mach-Zehnder interferometers have been proposed theoretically and implemented experimentally as tools to probe electric charges and statistics of anyons. The experimentally observed visibility of Aharonov-Bohm oscillations is maximal at a high transmission through an interferometer but simple theoretical expressions for the electric currents and noises are only available at low visibility. We consider an alternative version of a Mach-Zehnder interferometer, in which anyons tunnel between co-propagating chiral channels on the edges of quantum Hall liquids at the filling factors $n/(2n+1)$. We find simple exact solutions for any transmission. The solutions allow a straight-forward interpretation in terms of fractional charges and statistics.

\end{abstract}

\maketitle

A key concept in the field of topological matter is fractional statistics of excitations. It can be defined for extended objects \cite{3D-1,3D-2} in 3D and point anyons in two \cite{feldman:2021,LM:1977,Wilczek:1982,Halperin:1984} and sometimes three dimensions \cite{3D-non-Ab}. 

So far the evidence of topological systems with fractional statistics has been limited to 2D. Anyonic statistics were proposed in putative topological superconductors \cite{topsup} and RuCl$_3$, Ref. \onlinecite{RuCl3}, but the physics of those materials remains controversial. Very recently, evidence of a fractional Chern insulator state in twisted MoTe$_2$ has been reported \cite{MoTe-1,MoTe-2}. The bulk of research on topologically ordered materials has focused on the fractional quantum Hall effect \cite{HJ-Book} (FQHE).

\begin{figure}
    \centering
    \includegraphics[scale=1.3]{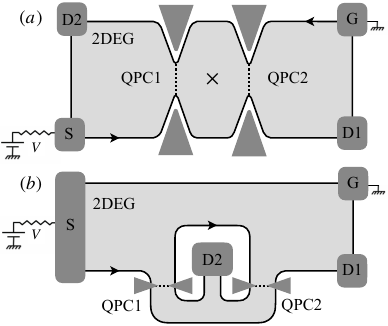}
    \caption{Schematics of the (a) Fabry-P{\'e}rot and (b) Mach-Zehnder interferometers. Quantum point contacts (QPC) bring the edges near to facilitate tunneling. Current flows into the device through  Ohmic contact S. (a) Fabry-P{\'e}rot interferometer involves tunneling between two contra-propagating edge modes. The current measured at drain D2 involves paths that braid around a localized anyon (marked by $\times$). Therefore, it is sensitive to anyon statistics. (b) In a Mach-Zehnder interferometer, two co-propagating modes are connected by the tunneling contacts. A common implementation requires the placement of an Ohmic contact (D2) inside the interference loop. Hence, each tunneling event changes the localized topological charge in the device. }
    \label{fig:1}
\end{figure}

Several experiments produced evidence of anyonic statistics in FQHE states. Non-Abelian statistics at the filling factor $\nu=5/2$ was demonstrated with the heat conductance technique \cite{therm1,therm2}. 
Anyon colliders \cite{col0,col1,col2,col3} have been used to probe Abelian statistics at $\nu=1/3$ and $2/5$. The most direct and intuitive approach to probing statistics consists in interferometry \cite{feldman:2021,SV:2019}. 
The idea is to split a beam of anyons into two beams on two sides of a localized anyon and measure the interference phase which depends on the mutual statistics of traveling and localized particles. This can be done in two ways. The Fabry-P{\'e}rot interferometry \cite{FP} involves two constrictions between two contra-propagating edge modes (Fig. \ref{fig:1}a). In a Mach-Zehnder interferometer \cite{MZ1,MZ2}, two co-propagating modes are connected by two tunneling contacts and an Ohmic contact is placed in its center (Fig. \ref{fig:1}b). There are also interesting multi-terminal versions of interferometry \cite{safi:2001,vishveshwara:2003,kim:2005,campagnano:2012}.

Despite early promising results \cite{Willett:2008,Willett:2010}, a convincing realization of interferometry proved challenging.
Evidence of fractional statistics in the simplest $\nu=1/3$ state from Fabry-P{\'e}rot interferometry \cite{Manfra:2020} arrived only in year 2020. A Mach-Zehnder interferometer
\cite{MZ:2023} in the same FQHE state was only realized in year 2023.
The difficulties were in part due to Coulomb effects \cite{Coulomb} and edge reconstruction \cite{reconstruction:2019}. Another issue consisted in the difficulty of theoretical analysis for strong tunneling between the two edges of the device.
Simple theoretical results are only available for weak tunneling, where the visibility of the Aharonov-Bohm oscillations is low. This is different from the integer quantum Hall effect (IQHE), where simple exact solutions exist for any visibility.

\begin{figure}
    \centering
    \includegraphics[scale=1.25]{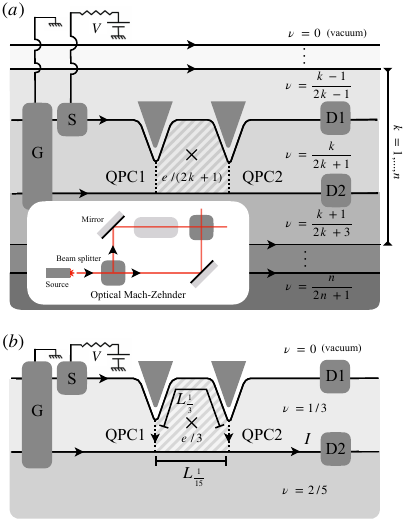}
    \caption{Schematics of the Mach-Zehnder geometry considered in this work. The edge of a $\nu=n/(2n+1)$ liquid contains $n$ co-propagating modes on either edge of a two-dimensional electron gas. On the right-moving edge, current flows into the device through Ohmic contact S to one of the edge modes. Two adjacent co-propagating edge modes are brought near to facilitate quantum tunneling. The localized anyon in the shaded region between the two constrictions is marked by $\times$. This geometry establishes an analogy with an optical Mach-Zehnder interferometer (inset) where, in contrast to the magnetic flux, the optical path length serves as a useful probe. (a) The edge mode, separating $\nu=(k-1)/(2k-1)$ and $\nu=k/(2k+1)$ incompressible states, is maintained at the potential bias $V$ with respect to the edge mode separating $\nu=k/(2k+1)$ and $\nu=(k+1)/(2k+3)$, where $k=1,\dots,n$. The tunneling anyon carries a charge $e/(2k+1)$. (b) We choose a representative system of $\nu=2/5$ where the two co-propagating edge modes $\phi_{\frac{1}{3}}$ and $\phi_{\frac{1}{15}}$ are separated by the $\nu=1/3$ incompressible liquid, facilitating anyon tunneling of charge $e/3$.}
    \label{fig:2}
\end{figure}

Here, we focus on another interferometer geometry that has been considered \cite{deviatov:2008} in IQHE, Fig. \ref{fig:2}. The geometry more directly parallels that of the original Mach-Zehnder interferometers studied in optics. As shown, two tunneling contacts are created between two co-propagating channels on the same edge. A similar geometry was implemented \cite{deviatov:2009,deviatov:2012} for the tunneling between the fractional edge modes separating $\nu=0$ from $\nu=1/3$ and $\nu=1/3$ from $\nu=2/3,~3/5,$ and $1$. This choice of FQHE modes results in essentially the same physics as in the standard Fabry-P{\'e}rot setup since the boundaries between $\nu=1/3$ and $\nu=2/3,~3/5$, and $1$ contain upstream modes. In this paper we observe that the geometry of Fig. \ref{fig:2} can be used to create a true Mach-Zehnder interferometer in FQHE with the tunneling between co-propagating modes and no Ohmic contacts inside the interferometer.  

A remarkable feature of this geometry is that it allows an easy exact solution for any visibility, including the experimentally optimal regime. Unlike most previous treatments of quantum Hall interferometers, we do not need to resort to perturbation theory. The solution is possible due to two key simplifications in comparison with the standard geometries. In contrast to the Fabry-P{\'e}rot setup, anyons cannot make multiple loops in the device. 
In contrast to the standard Mach-Zehnder setup, the localized topological charge in the interferometer remains fixed during an
experiment. Interestingly, the exact solution has essentially the same structure as in IQHE and contains information about the fractional charge and statistics of anyons. Note that this interferometer is not expected to show Coulomb-dominated behavior \cite{Coulomb}.

We consider an edge of an FQHE liquid with the filling factor $\nu=n/(2n+1)$. The bulk FQHE liquid is a daughter state \cite{WenBook} of the liquid with $\nu=(n-1)/(2n-1)$. Hence, the edge can be understood as a collection of co-propagating modes separating filling factors $k/(2k+1)$ and $(k-1)/(2k-1)$, where $k=1,\ldots, n$.
Fig. \ref{fig:2}a illustrates an interferometer constructed from the channel separating $\nu=(k+1)/(2k+3)$ and 
$\nu=k/(2k+1)$, and the adjacent channel separating $\nu=k/(2k+1)$ and $\nu=(k-1)/(2k-1)$. The incompressible region between the channels supports quasiparticles of charge $e/(2k+1)$. Such quasiparticles can tunnel between the two channels at the two constrictions. The Hamiltonian is the sum of the chiral Luttinger liquid Hamiltonians for the two channels plus two operators describing quasiparticle tunneling at the constrictions. Crucially, the scaling dimension of those operators is 1 just like in an IQHE system, which allows electron tunneling between two chiral non-interacting Fermi gases.  This scaling dimension is responsible for the exact solubility of the model via fermionization.

While our results do not depend much on $k$, we will focus on the simplest case of the bulk filling factor $2/5$,
Fig. \ref{fig:2}b. There are two edge channels \cite{WenBook}, which separate $\nu=0$ from $\nu=1/3$ and $\nu=1/3$ from $\nu=2/5$. The charge of the tunneling anyon is $e/3$.

The action is the sum of three contributions,
\begin{equation}
\label{1mzfp}
\mathcal{A}=\int dx dt \mathcal{L}_e-\int dt (T_1+T_1^\dagger)- \int dt (T_2+T_2^\dagger),
\end{equation}
where $T_1$ and $T_2$ describe anyon tunneling at the two constrictions and $\mathcal{L}_e$ is the edge Lagrangian density,
\begin{align}
\label{2mzfp}
\mathcal{L}_e=&-\frac{\hbar}{4\pi}\Big[3\partial_t\phi_{1/3}\partial_x\phi_{1/3}+3v_{1/3}(\partial_x\phi_{1/3})^2 \nonumber\\
&+15\partial_t\phi_{1/15}\partial_x\phi_{1/15}+15v_{1/15}(\partial_x\phi_{1/15})^2\Big],
\end{align}
with the two Bose-fields $\phi_{1/3}$ and $\phi_{1/15}$ describing the charge density on the outer and inner edge modes respectively: $\rho_{1/3}=e\partial_x\phi_{1/3}/2\pi$ and
$\rho_{1/15}=e\partial_x\phi_{1/15}/2\pi$. The edge-mode velocities are $v_{1/3}$ and $v_{1/15}$.
The operators 
\begin{eqnarray}
\label{3mzfp}
T_1&=&\Gamma_1\exp\left[i\phi_{1/3}(0)-5i\phi_{1/15}(0)\right]; \\
\label{4mzfp}
T_2&=&\Gamma_2\exp\left[i\phi_{1/3}(L_{1/3})-5i\phi_{1/15}(L_{1/15})\right], 
\end{eqnarray}
transfer an $e/3$-quasiparticle from the outer to inner edge. The amplitudes $\Gamma_i$ contain information about the charge and statistics of the tunneling quasiparticle. We denote the phase of  $\Gamma_1$ as $\phi$. 
The phase of $\Gamma_2=|\Gamma_2|e^{i\alpha}$
can be represented as the sum $\alpha=\phi+\alpha_0+\alpha_{\text{AB}}+\alpha_{\text{s}}$ of some non-universal phase $\phi+\alpha_0$, the Aharonov-Bohm phase $\alpha_{\text{AB}}$, and the statistical phase $\alpha_{\text{s}}$.
The contribution $\phi+\alpha_0$ is determined by microscopic details and does not depend on the magnetic field. The phase $\alpha_{\text{AB}}$ is proportional to the product of the anyon charge $e/3$ and the magnetic flux through the area between the two channels, shaded area in Fig. \ref{fig:2}b. The statistical phase $\phi_{\text{s}}=2\pi N_{\text{a}}/3$, where $N_{\text{a}}$ is the number of anyons localized inside the interferometer. This phase jumps when a new anyon enters the device in response to a change of the magnetic field. We assume that the left constriction has coordinate $x=0$ in both edge channels. In general, the lengths of the two channels between the constrictions are different. That's why the fields $\phi_{1/3}$ and $\phi_{1/15}$ are taken at different values of $x$ in the definition of $T_2$.

The electric current operator is defined as the commutator of the tunneling operator $T_1+T_2+{\rm h.~c.}$
with one-half times the charge difference between the two modes. Hence,
\begin{equation}
\label{5mzfp}
I=-i\frac{e}{3\hbar}[T_1^\dagger-T_1+T_2^\dagger-T_2].
\end{equation}
We will also need the correlation functions of the tunneling operators \cite{WenBook},
\begin{align}
&\Big\langle\exp\left[i\phi_{1/3}\left(t,a_{1/3}+b_{1/3}\right)-5i\phi_{1/15}\left(t,a_{1/15}+b_{1/15}\right)\right] \nonumber\\
&~~~~~~~\times\exp\left[-i\phi_{1/3}\left(0,a_{1/3}\right)+5i\phi_{1/15}\left(0,a_{1/15}\right)\right]\Big\rangle = \nonumber\\
&\frac{(\pi T/\hbar)^2}{\sin^{\frac{1}{3}}\left[\delta+i\frac{\pi T}{\hbar}\left(t-\frac{b_{1/3}}{v_{1/3}}\right)\right]
\sin^{\frac{5}{3}}\left[\delta+i\frac{\pi T}{\hbar}\left(t-\frac{b_{1/15}}{v_{1/15}}\right)\right]}.
\label{6mzfp}
\end{align}
The exponents in the above expression add up to two. This reflects the scaling dimension of one for $T_{1,2}$. The operators $T_{1,2}$ commute as locality demands. Note a difference from the usual Mach-Zehnder setup where locality requires Klein factors in tunneling operators.

The electrical potential difference between the channels is conveniently described in the interaction representation \cite{MZ2} by changing $\Gamma_{1,2} \rightarrow \Gamma_{1,2}\exp(-ieVt/3\hbar)$.

As a warming-up exercise, we consider the case of a single constriction, $\Gamma_2=0$. By rescaling the 
$x$-coordinate for each channel we can make the edge velocities equal. Let us rescale the coordinates so that the edge velocities become $u$. Let us next introduce new Bose fields in place of $\phi_{1/3}$ and
$\phi_{1/15}$:
\begin{eqnarray}
\label{7mzfp}
\phi_1&=&\frac{\sqrt{5}+1}{2}\phi_{1/3}-\frac{5-\sqrt{5}}{2}\phi_{1/15};  \\
\label{8mzfp}
\phi_2&=&\frac{\sqrt{5}-1}{2}\phi_{1/3}+\frac{5+\sqrt{5}}{2}\phi_{1/15}.  
\end{eqnarray}
The action becomes
\begin{align}
\label{9mzfp}
\mathcal{A}=&-\frac{\hbar}{4\pi}\int dtdx\sum_{k=1}^2\partial_x(\partial_t+u\partial_x)\phi_k \nonumber \\
&~-\int dt \left[\Gamma_1\exp\left(i\phi_1-i\phi_2\right)+{\rm h.c.}\right]. 
\end{align}
This action can be fermionized \cite{fermionization} in terms of Fermi operators $\psi_k=\exp(i\phi_k)$, where we ignore a dimensional normalization constant. Such a constant enters the relation between the tunneling amplitudes $\Gamma_i$ and the observable transmission probabilities. We will ignore the normalization constant below since our primary goal consists in connecting the transmission probabilities of the two individual constrictions with the transmission of the interferometer.   We rewrite (\ref{9mzfp}) as a free fermion action:
\begin{equation}
\label{10mzfp}
\mathcal{A}=i\hbar\int dt dx\sum_{k=1}^2 \psi_k^\dagger(\partial_t+u\partial_x)\psi_k
-\int dt [\Gamma_1 \psi^\dagger_2(0)\psi_1(0) +{\rm h.c.}].
\end{equation}
This model can be interpreted as the edge of an IQHE system at $\nu=2$. One just needs to remember that
the physical current (\ref{5mzfp}) is three times less than the current in the IQHE model and that the physical voltage $V$ is three times higher than the effective voltage in the IQHE model. Hence, the physical current is $1/3$ of the tunneling current in the model (\ref{10mzfp}) evaluated at $V/3$.

We will solve the model with the equation of motion approach. To avoid a delta-function of the coordinate in the equation of motion, we rewrite the tunneling term as an integral over a small vicinity of the origin: 
$\psi^\dagger_2(0)\psi_1(0)\rightarrow \frac{1}{\epsilon}\int_{0}^{\epsilon}dx\psi^\dagger_2(x)\psi_1(x)$,
where $\epsilon\rightarrow 0$. The equations of motion for the mode $\psi_{1,2}(x,t)=\psi_{1,2}(x)\exp(-iEt/\hbar)$ of energy $E=uk\hbar$ are
\begin{eqnarray}
\label{11mzfp}
(k+i\partial_x)\psi_1&=&\frac{\Gamma_1^*}{\epsilon u\hbar}\theta(x[\epsilon-x])\psi_2; \\
\label{12mzfp}
(k+i\partial_x)\psi_2&=&\frac{\Gamma_1}{\epsilon u\hbar}\theta(x[\epsilon-x])\psi_1. 
\end{eqnarray}
We find two independent solutions 
\begin{align}
\label{13mzfp}
\text{a)}\hspace{0.5cm}\psi_1(x)=e^{ikx} ~~~&{\text{and}}&\psi_2(x)=0~~~~~~~(x<0), \nonumber\\
 \psi_1(x)=t_1e^{ikx}~&{\text{and}}&\psi_2(x)=r_1e^{ikx}~ (x>0); 
\end{align}
\begin{align}
\label{14mzfp}
\text{b)}\hspace{0.3cm}\psi_1(x)=0 ~~~~~~~~~&{\text{and}}&\psi_2(x)=e^{ikx}~~~(x<0), \nonumber\\
\psi_1 (x)=-r_1^*e^{ikx}~&{\text{and}}&\psi_2(x)=t_1e^{ikx}~(x>0), 
\end{align}
where the transmission and reflection amplitudes between the two co-propagating channels are $t_1=\cos(|\Gamma_1|/u\hbar)$ and $r_1=-i\exp(i\phi)\sin(|\Gamma_1|/u\hbar)$ respectively.
The current can be computed from the energy-independent reflection $|r_1|^2$ into the parallel co-propagating channel  for fermions. It is temperature-independent and equals $I=e^2|r_1|^2V/9h$. The maximal tunneling conductance of $e^2/9h$ is achieved at $|r_1|=1$. Interestingly, it is greater than the conductance $e^2/15h$ of the inner channel. This happens due to Andreev reflection as discussed in Ref. \onlinecite{Andreev}.

We now map the problem with two constrictions onto free fermions. There is an essential difference from the single constriction case.
Indeed, in that case, the relevant correlation functions of the tunneling operators always have $b_{1/3}=b_{1/15}=0$ in Eq. (\ref{6mzfp}). As a result, the correlation functions are precisely the same in our problem and for the tunneling operator of free fermions. With two constrictions this is not necessarily the case. Fermionization only works if either the travel times between the two constrictions along the two edge channels are equal, $L_{1/3}/v_{1/3}=L_{1/15}/v_{1/15}$, or if the difference of the travel times is much less than the thermal and voltage lengths $\sim T^{-1}$ and $\sim (eV)^{-1}$ respectively. The latter condition is always satisfied if the interferometer is shorter than the thermal and voltage lengths but it can also work for an arbitrary large interferometer. 

We thus focus on the regime where fermionization applies. Besides the transmission and reflection amplitudes $t_1$ and $r_1$ for the first constriction, we define the transmission and reflection amplitudes $t_2$ and $r_2$ for the second constriction. The absolute values of $t_{1,2}$ and $r_{1,2}$ can be found experimentally by measuring transport through a single constriction when the second constriction is open. Probability conservation demands that $|t_i|^2+|r_i|^2=1$ and the absolute values of the transmission and reflection amplitudes are between 0 and 1. The relative phase of $r_1$ and $t_1$ is $-i\exp(i\phi)$ as discussed above. Similarly, the relative phase of $t_2$ and $r_2$ is 
$-i\exp(i\alpha)=-i\exp(i[\phi+\alpha_0+\alpha_{\text{AB}}+\alpha_{\text{s}}])$.
The total reflection amplitude for free fermions is $r=r_2t_1+r_1t_2$. This yields the following current between the inner and outer edges of the interferometer in the FQHE regime:
\begin{align}
\label{15mzfp}
I&=\frac{e^2V}{9h}|r_2t_1+r_1t_2|^2 \\
&=\frac{e^2V}{9h}\left[|r_1 t_2|^2+|r_2t_1|^2+2|r_1r_2t_1t_2|\cos(\alpha_0+\alpha_{\text{AB}}+\alpha_{\text{s}})\right]. \nonumber 
\end{align}
A change $\Delta\Phi$ of the magnetic flux between the edge channels results in the change $\Delta\alpha_{\text{AB}}=2\pi\Delta\Phi/3\Phi_0$, where $\Phi_0$ is a flux quantum. As the magnetic flux changes, quasiparticles or quasiholes enter the device, and $\alpha_{\text{s}}$ jumps by $2\pi/3$.

We now turn to the electric noise, for which an exact result can also be obtained. 
The zero frequency noise is defined \cite{noise-rev} as 
\begin{equation}
\label{16mzfp}
S=\int_{-\infty}^\infty dt [\langle I_{\text{D}}(0)I_{\text{D}}(t)+I_{\text{D}}(t)I_{\text{D}}(0)\rangle-2\langle I_{\text{D}}\rangle^2],
\end{equation}
where $I_{\text{D}}$ is the current in drain D1 or D2 and angular brackets denote the average. 
Our starting point is a general equation \cite{FH:2017} for the noise in chiral systems with tunneling
\begin{equation}
\label{17mzfp}
S=S_{\text{T}}-4T\frac{\partial \langle I\rangle}{\partial V}+4GT,
\end{equation}
where $S_{\text{T}}=\int_{-\infty}^\infty dt [\langle I(0)I(t)+I(t)I(0)\rangle-2\langle I\rangle^2]$ is the noise of the
tunneling current $I$, Eq. (\ref{5mzfp}), and $G$ is $e^2/3h$ or $e^2/15h$ for the outer or inner channel respectively.
The same equation applies to the noise in any of the two channels in the free electron model (\ref{10mzfp}), where the tunneling current operator is $3I(t)$ and $G=G_0=e^2/h$. We also should remember that the voltages differ by a factor of 3 in the FQHE and free fermion problems mapped onto each other.

An exact solution \cite{HF:2020} is available for the noise $S=S_{\text{F}}$ in the free fermion problem:
\begin{equation}
\label{18mzfp}
S_{\text{F}}=2eVG_0|r|^2(1-|r|^2)\left[\coth\left(\frac{eV}{2T}\right)-\frac{2T}{eV}\right] + 4G_0T,
\end{equation}
where $r=r_2t_1+r_1t_2$ is the total reflection amplitude. We now use equation (\ref{17mzfp}) to compute $S_{\text{T}}=S_{\text{TF}}$ for free fermions:
\begin{equation}
\label{19mzfp}
S_{\text{TF}}(V)=2eVG_0|r|^2(1-|r|^2)\left[\coth\left(\frac{eV}{2T}\right)-\frac{2T}{eV}\right]+4T|r|^2G_0.
\end{equation}
Finally, we compute the noise $S$ in the original model
\begin{align}
\label{20mzfp}
S=&\frac{1}{9}S_{\text{TF}}\left(V/3\right)-\frac{4G_0T|r|^2}{9}+4GT \\
=&\frac{2eV}{27}G_0|r|^2(1-|r|^2)\left[\coth\left(\frac{eV}{6T}\right)-\frac{6T}{eV}\right] + 4GT. \nonumber
\end{align}
The flux dependence enters through the coefficient 
\begin{align}
\label{21mzfp}
|r|^2(1-|r|^2)=&~R_0(1-R_0)-2{R_1^2}\nonumber \\
&+ 2R_1(1-2R_0)\cos(\alpha_0+\alpha_{\text{AB}}+\alpha_{\text{s}})\nonumber \\
&-2{R_1^2}\cos(2[\alpha_0+\alpha_{\text{AB}}+\alpha_{\text{s}}]),
\end{align}
where $R_0=|r_1t_2|^2+|r_2t_1|^2$ and $R_1=|r_1t_1r_2t_2|$. Thus, there are exactly two non-zero harmonics in the flux dependence. 
Only the second harmonics survives at $|t_1|=|r_1|=|t_2|=|r_2|=1/\sqrt{2}$.

In summary, we propose an anyonic Mach-Zehnder interferometer with two co-propagating edge channels and no Ohmic contact inside the device. The topological charge inside the device does not change after each tunneling event. Anyons, traveling through the device, cannot make multiple loops around localized particles. These properties open a way for a simple exact solution for the electric current and noise. The magnetic-field dependencies of the current and noise contain information about fractional charge and fractional statistics.

\begin{acknowledgments}
We thank M. Heiblum for useful discussions.
The research by NB, ZW, and DEF was supported in part by the National Science Foundation under Grant No. DMR-2204635. The collaboration of DEF and SV was supported in part by the National Science Foundation under Grant No. PHY-1748958.
\end{acknowledgments}


\begin{thebibliography}{99}

\bibitem{3D-1}
C. Aneziris, A. P. Balachandran, L. Kauffman, and A.
M. Srivastava, Novel statistics for strings and string ``Chern-Simon'' term,
\href{https://www.worldscientific.com/doi/10.1142/S0217751X91001210}{Int. J. Mod. Phys. A {\bf 6}, 2519 (1991)}.

\bibitem{3D-2} M. G. Alford, K.-M. Lee, J. March-Russell, and J.
Preskill, Quantum field theory of non-Abelian strings and vortices,
\href{https://www.sciencedirect.com/science/article/abs/pii/055032139290468Q?via}
{Nucl. Phys. B {\bf 384}, 251 (1992)}.

\bibitem{feldman:2021}
D. E. Feldman and B. I. Halperin, Fractional charge and fractional statistics in the quantum Hall effects, \href{https://doi.org/10.1088/1361-6633/ac03aa}{Rep. Prog. Phys. \textbf{84}, 076501 (2021)}.



\bibitem{LM:1977} J. M. Leinaas and J. Myrheim  On the theory of identical particles, 
\href{https://link.springer.com/article/10.1007/BF02727953}
{Nuovo Cimento B {\bf 37}, 1
(1977)}.

\bibitem{Wilczek:1982}
F. Wilczek, Quantum mechanics of fractional-spin particles, 
\href{https://journals.aps.org/prl/abstract/10.1103/PhysRevLett.49.957}
{Phys. Rev. Lett. {\bf 49}, 957 (1982)}.

\bibitem{Halperin:1984}
B. I. Halperin, Statistics of quasiparticles and the hierarchy of fractional quantized Hall states,
\href{https://journals.aps.org/prl/abstract/10.1103/PhysRevLett.52.1583}
{Phys. Rev. Lett. {\bf 52}, 1583 (1984)}.

\bibitem{3D-non-Ab}
J. C. Y. Teo and C. L. Kane,
Majorana fermions and non-Abelian statistics in three dimensions,
\href{https://journals.aps.org/prl/abstract/10.1103/PhysRevLett.104.046401}{Phys. Rev. Lett. {\bf 104}, 046401 (2010)}.

\bibitem{topsup}
M. Sato and Y. Ando, Topological superconductors: a review,
\href{https://iopscience.iop.org/article/10.1088/1361-6633/aa6ac7}
{Rep. Prog. Phys. {\bf 80}, 076501 (2017)}.

\bibitem{RuCl3}
P. A. Lee, Quantized (or not quantized) thermal Hall effect
and oscillations in the thermal conductivity in
the Kitaev spin liquid candidate $\alpha$-RuCl$_3$,
\href{https://www.condmatjclub.org/?p=4522}
{DOI: 10.36471/JCCM\_November\_2021\_02}.

\bibitem{MoTe-1}
C. Wang, X.-W. Zhang, X. Liu, Y. He, X. Xu, Y. Ran, T. Cao, and D. Xiao,
Fractional Chern insulator in twisted bilayer MoTe$_2$,
\href{https://arxiv.org/abs/2304.11864}{arXiv:2304.11864}.

\bibitem{MoTe-2}
Y. Zeng, Z. Xia, K. Kang, J. Zhu, P. Knüppel, C. Vaswani, K. Watanabe, T. Taniguchi, K. F. Mak, and J. Shan, Integer and fractional Chern insulators in twisted bilayer 
MoTe$_2$,
\href{https://arxiv.org/abs/2305.00973v2}
{arXiv:2305.00973}.

\bibitem{HJ-Book}
B. I. Halperin and J. K. Jain, eds., {\it Fractional Quantum Hall Effects: New Developments}
(World Scientific, 2020).

\bibitem{therm1}
M. Banerjee, M. Heiblum, A. Rosenblatt, Y. Oreg, D. E.
Feldman, A. Stern, and V. Umansky, Observed quantization of anyonic heat flow, \href
  {https://doi.org/10.1038/nature22052}{Nature {\bf 545}, 75 (2017)}.

\bibitem{therm2}
M. Banerjee, M. Heiblum, Umansky, D. E. Feldman,
Y. Oreg, and A. Stern, Observation of half-integer thermal Hall conductance, \href
  {https://doi.org/10.1038/s41586-018-0184-1}
{Nature {\bf 559}, 205 (2018)}.

\bibitem{col0}
B. Rosenow, I. P. Levkivskyi, and B. I. Halperin,
Current correlations from a mesoscopic anyon collider,
\href{https://journals.aps.org/prl/abstract/10.1103/PhysRevLett.116.156802}
{Phys. Rev. Lett. {\bf 116}, 156802 (2016)}.

 \bibitem{col1}
H. Bartolomei, M. Kumar, R. Bisognin, A. Marguerite,
J.-M. Berroir, E. Bocquillon, B. Pla\c{c}ais, A. Cavanna,
Q. Dong, U. Gennser, Y. Jin, and G. F{\`e}ve, 
Fractional statistics in anyon collisions, 
\href {https://doi.org/10.1126/science.aaz5601}
{Science {\bf 368}, 173 (2020)}.

\bibitem{col2}
P. Glidic, O. Maillet, A. Aassime, C. Piquard, A. Cavanna, U. Gennser, Y. Jin, A. Anthore, and F. Pierre,
Cross-correlation investigation of anyon statistics in the  $\nu=1/3$ and $2/5$
fractional quantum Hall states, \href{https://journals.aps.org/prx/abstract/10.1103/PhysRevX.13.011030}
{Phys. Rev. X {\bf 13}, 011030 (2023)}.

\bibitem{col3}
M. Ruelle, E. Frigerio, J.-M. Berroir, B. Pla\c{c}ais, J. Rech, A. Cavanna, U. Gennser, Y. Jin, and G. F{\`e}ve,
Comparing fractional quantum Hall Laughlin and Jain topological orders with the anyon collider,
\href{https://journals.aps.org/prx/abstract/10.1103/PhysRevX.13.011031}
{Phys. Rev. X {\bf 13}, 011031 (2023)}.

\bibitem{SV:2019}
V. Subramanyan and S. Vishveshwara, Correlations, dynamics, and interferometry of anyons in the lowest Landau level,
\href{https://iopscience.iop.org/article/10.1088/1742-5468/ab3aef}
{ J. Stat. Mech. {\bf 2019}, 104003 (2019)}.

\bibitem{FP}
C. C. Chamon, D. E. Freed, S. A. Kivelson, S. L. Sondhi, and X.-G. Wen, Two point-contact interferometer for quantum Hall systems,
\href{https://doi.org/10.1103/PhysRevB.55.2331}
{Phys. Rev. B {\bf 55}, 2331 (1997)}.

\bibitem{MZ1}
Y. Ji, Y. Chung, D. Sprinzak, M. Heiblum, D. Mahalu, and H. Shtrikman,  An electronic Mach–Zehnder interferometer,
\href{https://www.nature.com/articles/nature01503}
{Nature {\bf 422}, 415 (2003)}.

\bibitem{MZ2}
K. T. Law, D. E. Feldman, and Y. Gefen, Electronic Mach–Zehnder interferometer as a tool to probe fractional statistics,
\href{https://journals.aps.org/prb/abstract/10.1103/PhysRevB.74.045319}
{Phys. Rev. B {\bf 74}, 045319 (2006)}.

\bibitem{safi:2001}
I. Safi, P. Devillard, and T. Martin, Partition noise and statistics in the fractional quantum Hall effect,
\href{https://journals.aps.org/prl/abstract/10.1103/PhysRevLett.86.4628}
{Phys. Rev. Lett. {\bf 86}, 4628 (2001)}.

\bibitem{vishveshwara:2003}
S. Vishveshwara, 
Revisiting the Hanbury Brown–Twiss setup for fractional statistics,
\href{https://journals.aps.org/prl/abstract/10.1103/PhysRevLett.91.196803}
{Phys. Rev. Lett. {\bf 91}, 196803 (2003)}.

\bibitem{kim:2005}
E.-A. Kim, M. Lawler, S. Vishveshwara, and E. Fradkin,
Signatures of fractional statistics in noise experiments in quantum Hall fluids,
\href{https://journals.aps.org/prl/abstract/10.1103/PhysRevLett.95.176402}
{Phys. Rev. Lett. {\bf 95}, 176402 (2005)}.

\bibitem{campagnano:2012}
G. Campagnano, O. Zilberberg, I. V. Gornyi, D. E. Feldman, A. C. Potter, and Y. Gefen,
Hanbury Brown–Twiss interference of anyons,
\href{https://journals.aps.org/prl/abstract/10.1103/PhysRevLett.109.106802}
{Phys. Rev. Lett. {\bf 109}, 106802 (2012)}.


\bibitem{Willett:2008}
R. L. Willett, L. N. Pfeiffer, and  K. W. West,  Measurement of filling factor $5/2$ quasiparticle interference with observation of charge $e/4$ and $e/2$ period oscillations, 
\href{https://www.pnas.org/doi/full/10.1073/pnas.0812599106}
{Proc. Natl. Acad. Sci. U.S.A. {\bf 106}, 8853 (2008)}.

\bibitem{Willett:2010}
R. L. Willett, L. N. Pfeiffer, and K. W. West, Alternation and interchange of $e/4$ and $e/2$ period interference oscillations consistent with filling factor $5/2$ non-Abelian quasiparticles, 
\href{https://journals.aps.org/prb/abstract/10.1103/PhysRevB.82.205301}
{Phys. Rev. B {\bf 82}, 205301 (2010)}.

\bibitem{Manfra:2020}
J. Nakamura, S. Liang, G. C. Gardner, and M. J. Manfra,  Direct observation of anyonic braiding statistics at the $\nu=1/3$ fractional quantum Hall state,
\href{https://www.nature.com/articles/s41567-020-1019-1}
{Nature Phys. {\bf 16}, 931 (2020)}.

\bibitem{MZ:2023}
H. K. Kundu, S. Biswas, N. Ofek, V. Umansky, and M. Heiblum,
Anyonic interference and braiding phase in a Mach-Zehnder interferometer,
\href{https://www.nature.com/articles/s41567-022-01899-z}
{Nature Phys. {\bf 19}, 515 (2023)}.

\bibitem{Coulomb}
B. I. Halperin, A. Stern, I. Neder, and B. Rosenow, Theory of the Fabry–P\'erot quantum Hall interferometer, 
\href{https://journals.aps.org/prb/abstract/10.1103/PhysRevB.83.155440}{Phys. Rev. B {\bf 83}, 155440 (2011)}.

\bibitem{reconstruction:2019}
R. Bhattacharyya, M. Banerjee, M. Heiblum, D. Mahalu, and V. Umansky,
Melting of Interference in the Fractional Quantum Hall Effect: Appearance of Neutral Modes,
\href{https://journals.aps.org/prl/abstract/10.1103/PhysRevLett.122.246801}
{Phys. Rev. Lett. {\bf 122}, 246801 (2019)}.

\bibitem{deviatov:2008}
E. V. Deviatov and A. Lorke, Experimental realization of a Fabry-Perot-type interferometer by copropagating edge states in the quantum Hall regime,
\href{https://journals.aps.org/prb/abstract/10.1103/PhysRevB.77.161302}
{Phys. Rev. B {\bf 77}, 161302(R) (2008)}.

\bibitem{deviatov:2009}
E. V. Deviatov, B. Marquardt, A. Lorke, G. Biasiol, and L. Sorba,
Interference effects in transport across a single incompressible strip at the edge of the fractional quantum Hall system, 
\href{https://journals.aps.org/prb/abstract/10.1103/PhysRevB.79.125312}
{Phys. Rev. B {\bf 79}, 125312 (2009)}.

\bibitem{deviatov:2012}
E. V. Deviatov, S. V. Egorov, G. Biasiol, and L. Sorba, Quantum Hall Mach–Zehnder interferometer at fractional filling factors,
\href{https://iopscience.iop.org/article/10.1209/0295-5075/100/67009}
{Europhys. Lett. {\bf 100}, 67009 (2012)}.

\bibitem{WenBook}
X.-G. Wen, \textit{Quantum Field Theory of Many-Body Systems: From the Origin of Sound to an Origin of Light and Electrons} (Oxford University Press, Oxford, 2004).

\bibitem{fermionization}
J. von Delft and H. Schoeller,
Bosonization for beginners - refermionization for experts,
\href{https://onlinelibrary.wiley.com/doi/10.1002/andp.19985100401}
{Annalen Phys. {\bf 7}, 225 (1998)}.

\bibitem{Andreev}
N. P. Sandler, C. de C. Chamon, and E. Fradkin,
Andreev reflection in the fractional quantum Hall effect,
\href{https://journals.aps.org/prb/abstract/10.1103/PhysRevB.57.12324}
{Phys. Rev. B {\bf 57}, 12324 (1998)}.

\bibitem{noise-rev}
Ya. M. Blanter and M. B{\"u}ttiker, Shot noise in mesoscopic conductors,
\href{https://doi.org/10.1016/S0370-1573(99)00123-4} 
{Phys. Rep. {\bf 336}, 1 (2000).}

\bibitem{FH:2017}
D. E. Feldman and M. Heiblum,
Why a noninteracting model works for shot noise in fractional charge experiments,
\href{https://doi.org/10.1103/PhysRevB.95.115308}{Phys. Rev. B {\bf 95}, 115308 (2017)}.

\bibitem{HF:2020}
M. Heiblum and D. E. Feldman, Edge probes of topological order,
\href{https://doi.org/10.1142/S0217751X20300094}{Int. J. Mod. Phys. A  {\bf 35}, 2030009 (2020)}.



\end{thebibliography}
\end{document}